\documentclass[useAMS,usenatbib]{mn2e}
\usepackage{times}

\title[Gravitationally lensed water masers]{A search for gravitationally lensed water masers in dusty quasars and star-forming galaxies}
\author[J. P. McKean et al.]{J. P. McKean,$^{1}$\thanks{E-mail: mckean@astron.nl} C. M. V. Impellizzeri,$^{2}$ A. L. Roy,$^{3}$ P. Castangia,$^{4}$ F. Samuel,$^{2}$
\newauthor A. Brunthaler,$^{3}$ C. Henkel$^{3}$ and O. Wucknitz$^{5}$\\
$^{1}$ASTRON, Oude Hoogeveensedijk 4, 7991 PD Dwingeloo, the Netherlands\\
$^{2}$National Radio Astronomy Observatory, 520 Edgemont Road, Charlottesville, VA 22903, USA\\
$^{3}$Max-Planck-Institut f\"{u}r Radioastronomie, Auf dem H\"{u}gel 69, D-53121 Bonn, Germany\\
$^{4}$INAF-Osservatorio Astronomico di Cagliari, Loc. Poggio dei Pini, Strada 54, I-09012 Capoterra (CA), Italy\\
$^{5}$Argelander-Institut f\"{u}r Astronomie, Auf dem H\"{u}gel 71, D-53121 Bonn, Germany}
\begin{document}

\date{Accepted 2010 September 1. Received 2010 September 1; in original form 2010 July 13}

\pagerange{\pageref{firstpage}--\pageref{lastpage}} \pubyear{2010}

\maketitle

\label{firstpage}

\begin{abstract}
Luminous extra-galactic water masers are known to be associated with active galactic nuclei and have provided accurate estimates for the mass of the central supermassive black hole and the size and structure of the circumnuclear accretion disk in nearby galaxies. To find water maser systems at much higher redshifts, we have begun a survey of known gravitationally lensed  quasars and star-forming galaxies. In this paper, we present a search for 22 GHz (rest frame) water masers toward five dusty, gravitationally lensed quasars and star-forming galaxies at redshifts between 2.3 and 2.9 with the Effelsberg radio telescope and the Expanded Very Large Array. Our observations do not find any new definite examples of high redshift water maser galaxies, suggesting that large reservoirs of dust and gas are not a sufficient condition for powerful water maser emission. However, we do find the tentative detection of a water maser system in the active galaxy IRAS 10214+4724 at redshift 2.285. Our survey has now doubled the number of gravitationally lensed galaxies and quasars that have been searched for high redshift water maser emission. We also present an updated analysis of the high redshift water maser luminosity function that is based on the results presented here and from the only cosmologically distant ($z>$~1) water maser galaxy found thus far, MG J0414+0534 at redshift 2.64. By comparing with the water maser luminosity function locally and at moderate redshifts, we find that there must be some evolution in the luminosity function of water maser galaxies at high redshifts. By assuming a moderate evolution [$(1+z)^4$] in the water maser luminosity function, we find that blind surveys for water maser galaxies are only worthwhile with extremely high sensitivity like that of the planned Square Kilometre Array (Phase 2), which is scheduled to be completed by 2020. However, instruments like the EVLA and MeerKAT will be capable of detecting water maser systems similar to the one found from MG J0414+0534 through dedicated pointed observations, providing suitable high-redshift targets can be selected.  
\end{abstract}


\section{Introduction}

The 6$_{16}-$5$_{23}$ transition of water (emitted rest-frequency of 22.23508 GHz) shows spectacularly luminous extra-galactic maser emission, with isotropic line luminosities of up to 23\,000\,$L_{\odot}$ (for a review see, e.g., \citealt{lo05}). The maser action results from collisional excitation of water molecules in extremely dense gas clouds and requires particle densities of 10$^7$ to 10$^{11}$~cm$^{-3}$ and gas temperatures $>$\,300\,K. All interferometric studies of powerful water masers ($>$\,10\,$L_{\odot}$) found thus far have always shown them to be associated with active galactic nuclei (AGN). High-resolution studies with very long baseline interferometry (VLBI) of $\sim$10 of these systems have shown that the water masers are associated with either the circumnuclear accretion disk within a few parsecs of the central supermassive black hole \citep{miyoshi95,herrnstein99,kondratko08,reid09} or within the inner part ($<$\,30~pc) of the relativistic radio jets that are ejected from some AGN \citep{claussen98,peck03}. The high surface brightness and proximity to the central engine make water maser systems a powerful tool for studying the properties of AGN. For example, water masers have been used to measure the mass of the central supermassive black hole \citep{miyoshi95,greenhill03a,reid09,Kuo10}, the size and structure of the circumnuclear accretion disk \citep{greenhill03a,argon07,humphreys08,kondratko08}, the shock speeds and densities of radio jets \citep{peck03}, and to determine accurate geometrical distances to the host galaxies \citep{herrnstein99,braatz10} -- all of which are competitive with the results obtained using other methods.

Surveys for water maser systems at low redshifts ($z\leq$\,0.06) have tended to focus on Seyfert 2 and LINER galaxies (e.g. \citealt{greenhill03b,braatz04}). This is because the expected edge-on orientation of the circumnuclear accretion disk for these galaxies should result in a larger path length of velocity-coherent molecular gas which is needed for the maser amplification. The success rate of these surveys is of order $\sim$5 per cent, with around $\sim$130 water maser galaxies currently known locally (e.g. \citealt{braatz07}; and unpublished). Surveys have also targeted nearby spiral and starburst galaxies with bright far-infrared (FIR) flux-denisties \citep{henkel05,castangia08,surcis09}. These surveys have found 15 water maser galaxies and have a detection rate of 23 per cent \citep{surcis09}. A snapshot survey of 611 luminous active and non-active galaxies in the local Universe ($v \leq$~5000\,km) was carried out by \citet{braatz08}. They found only 8 water maser systems (a detection rate of about $\sim$~1 per cent), which demonstrates the rarity of these objects. Finding water maser galaxies at higher redshifts has the potential to constrain the nature of dark energy through the accurate measurement of geometrical distances. However, due to the limited sensitivities and the small range of frequencies covered by current radio telescopes, surveys for high redshift water maser galaxies had yielded non-detections \citep{bennert09}, with the notable exception being the detection of the most luminous water maser system known in a type 2 quasar at redshift 0.66 \citep{barvainis05}. 

To overcome the sensitivity limitations we have started a search for water masers from known gravitationally lensed quasars. Observing only gravitationally lensed active galaxies has the advantage of pre-selecting a population of distant AGN ($z > 1$) and of using the magnification provided by the foreground gravitational lensing galaxy to increase the observed flux density of any water masers in the background AGN. Such lensing magnifications can range up to hundreds, depending on the alignment of the lens and the background source. The gravitational lensing technique has already been used successfully to study the interstellar medium and molecular gas content of high redshift quasars through the detection of, for example, carbon monoxide \citep{barvainis94}. We found a luminous water maser in the first gravitationally lensed quasar we observed, in the type 1 quasar MG J0414+0534 at redshift 2.64 \citep{impellizzeri08}, which is by far the most distant object to show water vapour emission. The maser line from MG J0414+0534 is broad with a full width at half maximum (FWHM) of $\sim$\,45 km\,s$^{-1}$, is blueshifted by $-$300~km\,s$^{-1}$ from the systemic velocity of the quasar and has an (unlensed) apparent isotropic luminosity of $\sim$\,8300~$L_{\odot}$. Our initial hypothesis, based on spectra obtained with the Expanded Very Large Array (EVLA) and the Effelsberg radio telescope, was that the maser originates in the relativistic jet of this quasar as it interacts with a molecular cloud lying close to the supermassive black hole. Follow-up VLBI and monitoring observations are currently being analyzed to investigate this hypothesis and will be presented in forthcoming papers.

We have now begun a survey of other distant gravitationally lensed quasars, and have also extended our search to lensed starbursting galaxies, using the Effelsberg radio telescope and the EVLA. The aim of this survey is to find additional detections of high redshift water masers, the results of which are presented in this paper. In the case of the only currently confirmed high redshift water maser galaxy, MG J0414+0534, we know that the source is highly dust reddened and hosts a powerful AGN that is emitting a relativistic jet. This suggests that luminous water masers that are associated with AGN are more likely to be found in high redshift galaxies with large amounts of dust and/or powerful radio jets. To investigate the first of these possible effects, we have limited the sample studied in this paper to include only radio-quiet quasars and star-forming galaxies with previously measured strong far-infrared (FIR) emission or molecular CO line emission.

For all calculations we adopt an $\Omega_{M} =$~0.3, $\Omega_{\Lambda}=$~0.7 spatially flat Universe, with a Hubble constant of $H_{0} =$~70~km\,s$^{-1}$~Mpc$^{-1}$, and a solar luminosity of $L_{\odot}=$~3.939~$\times$~10$^{26}$ W.

\section{Targets}
\label{targets}

We targeted and prioritized those gravitational lens systems with i) background AGN or star-forming galaxies at redshifts between 2.29 and 2.87 (due to the bandpass of the receiver that was used), ii) had high lensing magnifications and iii) showed previous detections of molecular emission (e.g. CO, HCN, etc.) or a large FIR luminosity. We also limited our sample to those objects that were observable with the Effelsberg radio telescope and the EVLA. The resulting sample is not statistically complete.

\subsection{RX J0911+0551}

RX J0911+0551 is comprised of four images of an X-ray detected quasar at redshift 2.80 \citep{bade97,burud98}, which is gravitationally lensed by a group of galaxies at redshift 0.769 \citep{kneib00}. The total magnification of the quasar is $\sim$\,22. CO (3--2) has been tentatively detected  by \citet{hainline04} and the large sub-mm luminosity of the lensed quasar suggests a cold dust mass of $\sim$10$^{8}\,M_{\odot}$ \citep{barvainis02}.

\subsection{IRAS 10214+4724}

IRAS~10214+4724 is a highly magnified and dust obscured gravitationally lensed quasar at redshift 2.285 \citep{rowan-robinson91}. This system has four lensed images that are formed by a massive foreground lensing galaxy at redshift 0.893 \citep*{lacy98}. The magnification of the three merging lensed images is $\la$\,100 (e.g. \citealt{broadhurst95}). The lensed source is extremely luminous in the FIR \citep{rowan-robinson91} and sub-mm \citep{barvainis02}, and optical polarimetry found evidence of scattered light (\citealt{lawrence93}). A large quantity of molecular gas from CO \citep{brown91,brown92,solomon92,downes95}, HCN \citep{vandenbout04} and neutral carbon \citep{weiss05a} has also been reported. Finally, near-infrared imaging spectroscopy found evidence for both circumnuclear star formation and an AGN from narrow (but spatially extended) and broad H$\alpha$ emission \citep{kroker96}. All of these data are consistent with the central engine of the AGN being obscured and there being nuclear driven star-formation in a region rich in molecular gas.

\subsection{H~1413+117 -- The Cloverleaf}

H~1413+117 has four lensed images of a radio-quiet quasar at redshift 2.560 that are in the shape of a cloverleaf \citep{magain88}. The optical emission from the four lensed images is so strong that attempts to measure the redshift of the lens have not been successful. The total magnification of the background quasar is estimated to be $\sim$\,10 \citep{kayser90}. The system has a plethora of molecular emission and absorption detections; for example, CO (3--2; \citealt{barvainis94}, 4--3, 5--4, 7--6; \citealt{barvainis97} and 5--6, 8--7, 9--8; \citealt{bradford09}),  HCN (1--0; \citealt{soloman03}) and CN (3--2; \citealt{riechers07}). The quasar is also ultra-luminous in the sub-mm and has an estimated dust mass of $\sim$\,4\,$\times$\,10$^{8}\,M_{\odot}$ \citep{barvainis02}. Previous observations searching for maser activity failed to detect any water maser emission from this quasar down to an unlensed isotropic luminosity limit of 4600~$L_{\odot}$, with a 10~km\,s$^{-1}$ spectral resolution \citep{wilner99}.

\subsection{MS 1512--cB58}

MS 1512$-$cB58 is a star-forming Lyman break galaxy at redshift 2.727 which is gravitationally lensed by the massive cluster MS 1512+36 at redshift 0.37 \citep{yee96}. The galaxy is lensed into a gravitational arc by the cluster and is estimated to have a high total magnification of $\ga$\,50 \citep{seitz98}. CO (3--2) has been detected from the galaxy \citep{baker04}, which along with the large potential maser magnification led to MS 1512$-$cB58 being included in our sample.

\subsection{SMM J16359+6612}

SMM J16359+6612 is composed of three lensed images of a pair of merging galaxies at redshift 2.516. The lens is the massive cluster A2218 at redshift 0.18, and provides a total magnification of $\sim$\,45 \citep{kneib04}. The merging lensed galaxies show strong sub-mm emission \citep{kneib04} and a number of CO detections (transitions 3--2, 4--3, 5--4, 6--5, 7--6; e.g., \citealt{weiss05b,kneib05}). SMM J16359+6612 was recently observed for water maser emission with the EVLA by \citet{edmonds09} and  \citet{wagg09}. However, no water vapour was detected down to an unlensed luminosity of $\sim$\,5400\,$L_{\odot}$ within a 10 km\,s$^{-1}$ channel.

\section{Observations}
\label{observations}

In this section, our observations with the Effelsberg 100 m radio telescope and subsequent follow-up observations with the EVLA of tentative water maser detections are presented.

\subsection{Effelsberg observations and analysis}

The spectroscopic observations with the Effelsberg 100 m radio telescope were carried out with the 5 cm dual-polarization HEMT receiver at the primary focus. This receiver is sensitive between 5.75 and 6.75 GHz, which for the detection of 22.2~GHz water masers corresponds to a redshift range of 2.29 to 2.87. The primary beam of the telescope at these frequencies has a FWHM of $\sim$120~arcsec, which is easily large enough to encompass the angular extent of the gravitationally lensed images.

The data were taken using the position-switching mode where a 2.5~min on-source scan was immediately followed by a 2.5~min off-source scan. This off-source scan was subtracted from the on-source scan to remove the contribution of the sky background and the instrument from the on-source measurements. Observations of standard flux-density calibrators were also taken to determine the antenna gain (3C\,295; \citealt{ott94}). The spectra were formed using the 16\,384-channel Fast Fourier Transform Spectrometer (FFTS; \citealt{klein06}). Both the 20 and 100 MHz bandwidths that are available with the FFTS were used during these observations. A summary of the integration times, observing dates and bandwidths used for each lens system is given in Table \ref{obs-log}.

The spectra were analysed using the {\sc class77} package within {\sc gildas}. Each scan was inspected for any strong gain variations or radio frequency interference (RFI), with any spurious scans or frequency ranges flagged. The scans were then averaged to form a spectrum of the lens system for each observing epoch. A low-order polynomial was then fitted to the spectrum to subtract any continuum emission from the lensed images. For those lens systems that were observed on more than one occasion, an average spectrum weighted by the integration times was also produced. We smoothed the spectra using Hanning smoothing to a spectral resolution of 10~km\,s$^{-1}$~channel$^{-1}$. Where we found a possible water maser line, a Gaussian line profile was fitted to the unsmoothed spectrum to determine the FWHM, the integrated line flux, the line peak flux density and the line velocity.

\begin{table*}
\begin{center}
\caption{Observing logs for the Effelsberg radio telescope observations of gravitationally lensed galaxies.}
\begin{tabular}{lccclrccc} \hline
Lens system		& RA           	& Dec		& Redshift		& Date 			& Time		& Centre frequency	& Velocity range	& Bandwidth	\\
       				& (J2000)     	& (J2000)		&			& 				& (min)		& (GHz)			& (km\,s$^{-1}$)		& (MHz)	\\ \hline
RX~J0911+0551	& 09:11:27.500	& 05:50:52.00	& 2.793		& 2008 July 04		& 152.5		& 5.8617			& $-$2018 : 2207	& 100	\\
IRAS 10214+4724	& 10:24:34.599	& 47:09:11.00	& 2.285		& 2007 November 12 & 209.1 		& 6.7678			& $-$267 : 252	& 20\\
				&			&			&			& 2008 March 14	& 116.1		&				& $-$313 : 304						& 20\\	
				&			&			&			& 2008 March 17	& 215.7		&				& $-$266 : 252						& 20\\
H~1413+117		& 14:15:46.300	& 11:29:43.00	& 2.561		& 2008 August 15	& 283.7		& 6.2440			& $-$1898 : 2113	& 100\\
MS~1512$-$cB58	& 15:14:22.219	& 36:36:24.79	& 2.720		& 2008 May 27		& 103.1		& 5.9772			& $-$2014 : 2216	& 100\\
				&			&			&			& 2008 August 15	& 87.8		&				& $-$1884 : 2154					& 100\\
SMM~J16359+6612	& 16:35:52.501	& 66:12:15.01	& 2.520		& 2008 May 26		& 332.5		& 6.3167			& $-$1846 : 2022	& 100\\
\hline
\end{tabular}
\label{obs-log}
\end{center}
\end{table*}

\subsection{Expanded Very Large Array observations}

Our spectroscopic observations found tentative detections of water maser lines from the gravitational lens system IRAS 10214+4724 (see Section \ref{results} for details). To  confirm these detections independently, we carried out interferometric spectroscopic observations with the EVLA. These observations would also spatially match any water maser emission with the lens system. This is the same strategy that was used to confirm the detection of a water maser from the lensed quasar MG J0414+0534 \citep{impellizzeri08}.

IRAS 10214+4724 was observed at a central frequency of 6.7675 GHz with the EVLA on 2008 May 30 and 31 using 15 antennas that had recently been upgraded with the new C-band receivers. The data were taken through a total bandwidth of 6.25 MHz and split into 64 channels. This gave a spectral coverage of 277~km\,s$^{-1}$ and a spectral resolution of 4.3~km\,s$^{-1}$ per channel. IRAS 10214+4724 was known to be a weak radio source \citep{lawrence93}, therefore the observations were phase-referenced by observing the calibrator JVAS\,J0958+474 every $\sim$\,15 minutes to determine the phase and amplitude solutions. Observations of another nearby continuum source, JVAS\,J1033+412, were also taken throughout the run to verify the calibration process. The flux-density calibration was determined with 3C\,147 and 3C\,286 assuming 6.7 GHz flux-densities of 5.858 and 6.051~Jy, respectively. IRAS 10214+4724 was observed for $\sim$~8.5~h in total. The data were reduced in the standard way using {\sc casa}. A data cube was created using all of the channels, except for the first and last five channels. The noise per channel was 0.57 mJy~beam$^{-1}$ and the noise in the resulting continuum map was 78~$\mu$Jy~beam$^{-1}$. 

\section{Results}
\label{results}

We now present a brief description of the radio spectra obtained for each gravitational lens system. Isotropic line luminosities, and upper limits in the case of non-detections, have been calculated using
\begin{equation}
\frac{L_{\rm H_{2}O}}{L_\odot} = \frac{1}{m} \, \frac{\rm 0.023}{1+z} \, \frac{\int S\, d\nu}{\mbox {[Jy km\,s$^{-1}$]}} \, \frac{D^2_L}{\mbox {[Mpc$^2$}]},
\label{lum}
\end{equation}
where $m$ is the estimated lensing magnification, $z$ is the redshift of the background source, $\int S\,d\nu$ is the integrated line profile, $D_L$ is the luminosity distance and $L_{\rm H_{2}O}$ is the isotropic luminosity. We quote luminosities after correcting for the estimated magnification of the lens. However, these values should be taken with caution since the actual magnification of any given maser region can be calculated only from high-resolution imaging of the masers (to determine their position relative to the lensing galaxy) and lens modelling. This is particularly important for those cases with extended lensed emission where there can be a large differential magnification over the extent of the lensed images. All of the luminosity upper limits are for rectangular maser lines of width 10 km\,s$^{-1}$.

\begin{figure*}
\begin{center}
\setlength{\unitlength}{1cm}
\begin{picture}(12,12.5)
\put(-4.2,14.0){\includegraphics{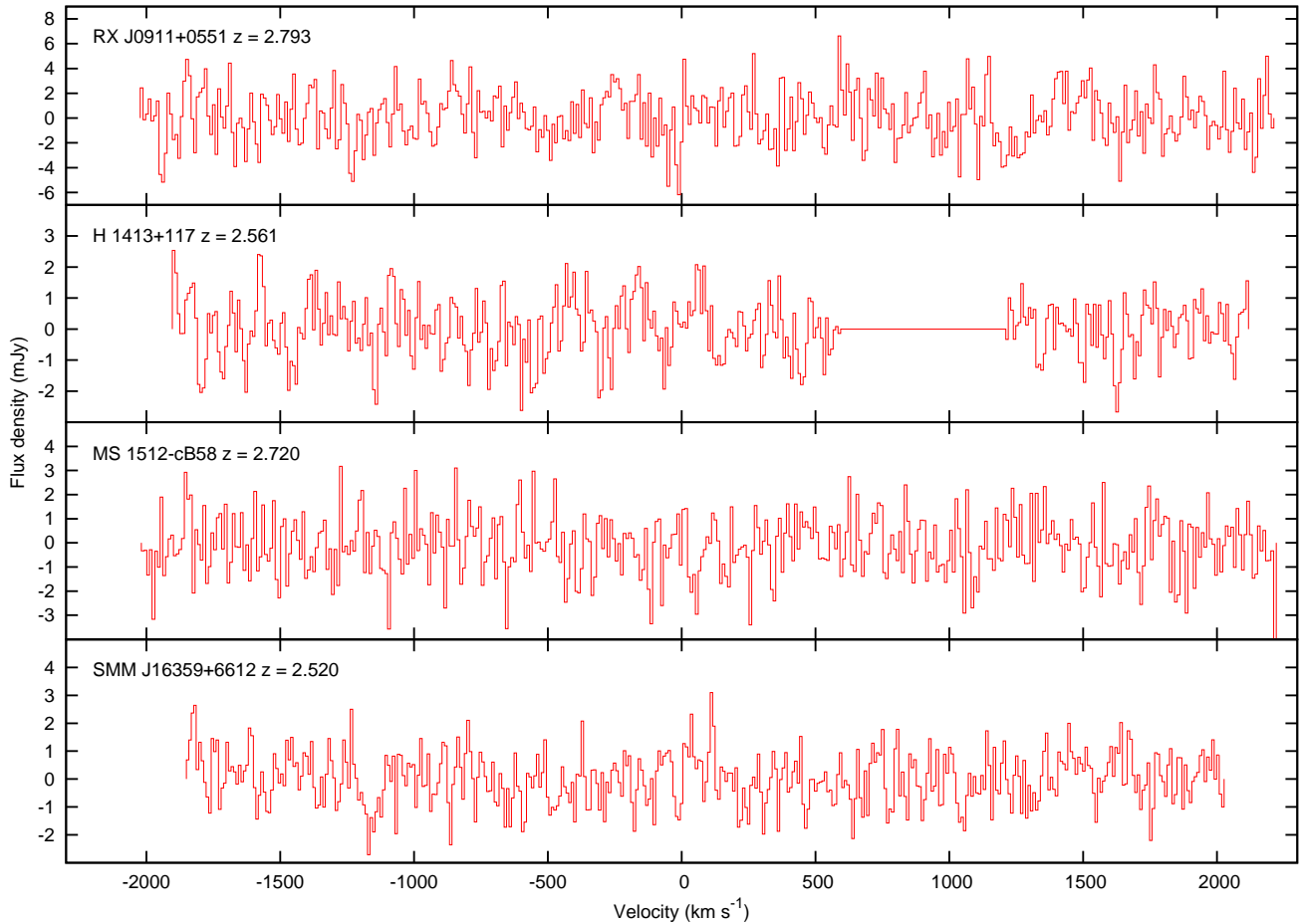}}
\end{picture}
\caption{The Effelsberg spectra of RX J0911+0551 ({\it top}), H 1413+116 ({\it top-middle}), MS 1512$-$cB58 ({\it bottom-middle}) and SMM J16359+6612 ({\it bottom}). The spectra have a resolution of 10 km\,s$^{-1}$ channel$^{-1}$ with an rms of 1.6, 0.9, 1.2 and 0.9 mJy~channel$^{-1}$, respectively for the four target lens systems. All velocities are relative to the systemic velocities of the quasars, and are in the heliocentric frame. The uncertainty in the optical systemic velocity of each target is around $\pm$\,100 km\,s$^{-1}$. These targets were observed using 100 MHz bandwidth.} 
\label{all-spectra}
\end{center}
\end{figure*}

\subsection{RX J0911+0551}

The spectrum of RX J0911+0551 is presented in Fig. \ref{all-spectra} and shows no definite detection of water maser emission within $\pm$\,2000 km\,s$^{-1}$ of the systemic velocity. The rms noise in the spectrum is 1.6 mJy for a 10 km\,s$^{-1}$ channel. This corresponds to a luminosity limit of $<$\,7200~$L_\odot$ (3\,$\sigma$) when corrected for the estimated magnification from the lens ($\sim$\,22).

\subsection{IRAS 10214+4724}

\begin{figure}
\begin{center}
\setlength{\unitlength}{1cm}
\begin{picture}(6,10.2)
\put(-2.4,11.7){\includegraphics{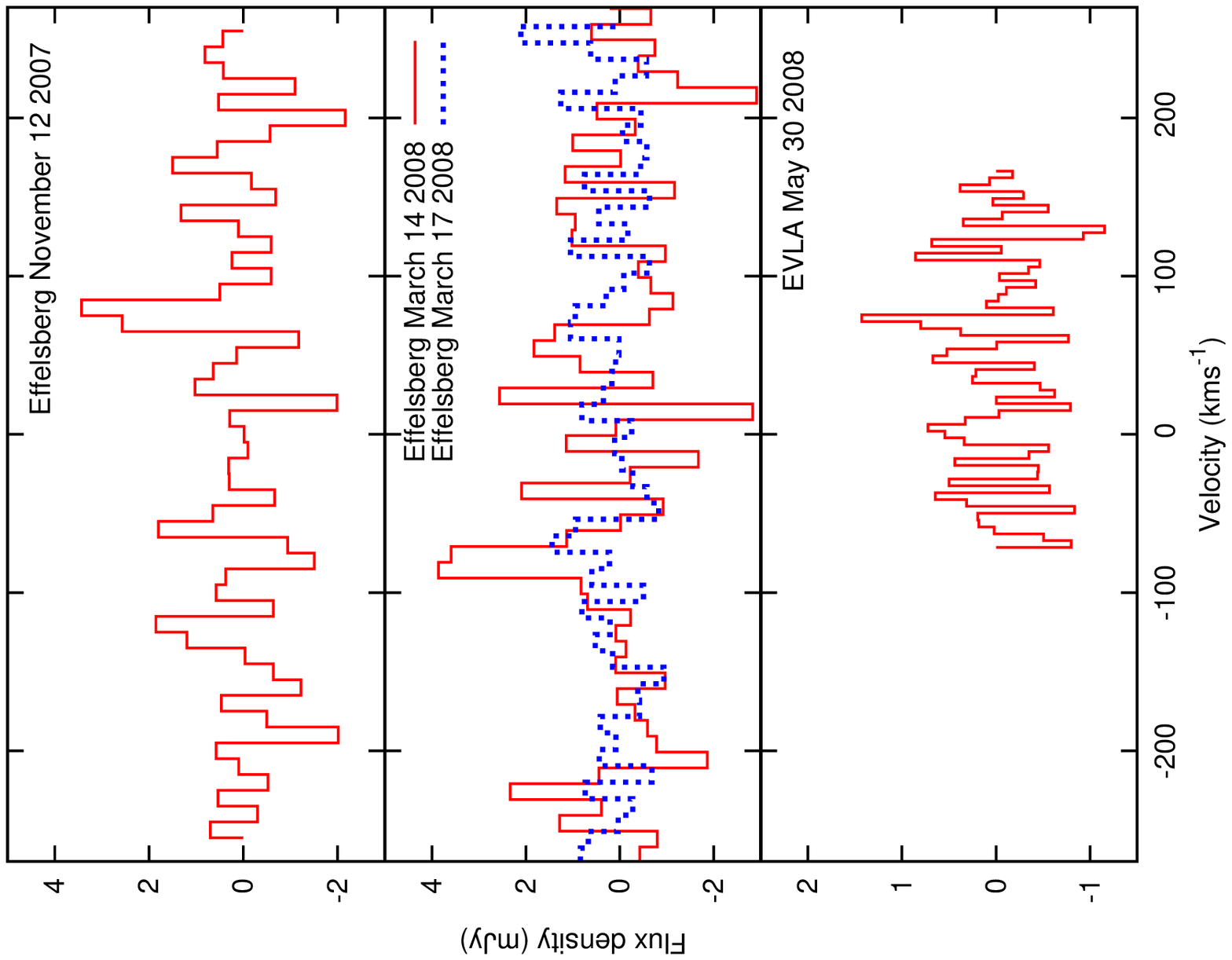}}
\end{picture}
\caption{The Effelsberg and EVLA spectra of IRAS 10214+4724. Top panel: the Effelsberg spectrum taken on 2007 November 12 shows a clear peak at $+$77~km\,s$^{-1}$ from systemic. The rms is 1.0~mJy~channel$^{-1}$ (10 km\,s$^{-1}$ channel width). Middle panel: the Effelsberg spectra taken on 2008 March 14 and 17. There is evidence of a peak at $-$81~km\,s$^{-1}$ on March 14, but nothing was found on March 17. The rms is 1.2 and 0.6~mJy~channel$^{-1}$, respectively. Bottom panel: the EVLA spectrum taken on 2008 May 30. Again, a faint emission line is detected at $\sim$~72~km\,s$^{-1}$. The rms is 0.6~mJy~channel$^{-1}$ (4.3 km\,s$^{-1}$ channel width). All velocities are relative to the systemic velocity of IRAS 10214+4724 at redshift 2.285, and are in the heliocentric frame. The bandwidth of the Effelsberg and EVLA observations was 20 and 6.25 MHz, respectively.} 
\label{IRAS10214-spectra}
\end{center}
\end{figure}

Our Effelsberg spectra of IRAS 10214+4724 are presented in Fig.~\ref{IRAS10214-spectra} and show two tentative detections of water maser emission. The first epoch of data that were taken with Effelsberg shows a clear 5\,$\sigma$ peak in emission at 77\,$\pm$\,1 km\,s$^{-1}$ relative to the systemic velocity of the quasar. The peak of the fitted Gaussian is 4.7 mJy and the Gaussian FWHM of the line is 13\,$\pm$\,3 km\,s$^{-1}$. The integrated total intensity of the potential maser line is 0.066~Jy~km\,s$^{-1}$, giving an isotropic luminosity of 3100~$L_\odot$ (for a lensing magnification of 50).  The noise in the spectrum within a 10~km\,s$^{-1}$ channel is 1.0~mJy~channel$^{-1}$. The follow-up observations of IRAS 10214+4724 with Effelsberg on 2008 March 14 and 15 did not detect this emission feature, but did show another emission line at $-$81\,$\pm$\,2 km\,s$^{-1}$ that was significant at 4\,$\sigma$. This line has a fitted Gaussian peak flux density of 4.9 mJy and a FWHM of 18\,$\pm$\,3 km\,s$^{-1}$. The integrated intensity of the line is 0.096~Jy~km\,s$^{-1}$ and the isotropic luminosity is 4500~$L_\odot$. However, this second emission feature was not detected on both days during the second Effelsberg observing epoch. The rms noise of the spectra taken on March 15 and 17 are 1.2 and 0.6~mJy~channel$^{-1}$, respectively, within 10~km\,s$^{-1}$ wide channels.

\begin{figure}
\begin{center}
\setlength{\unitlength}{1cm}
\begin{picture}(6,7.6)
\put(-1.2,-0.2){\includegraphics{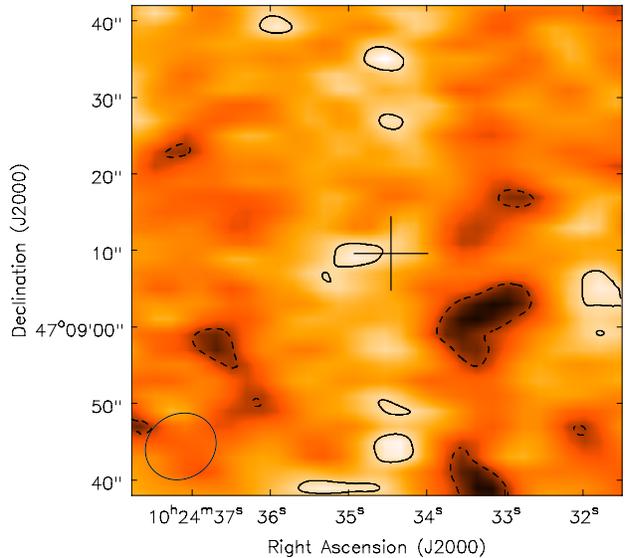}}
\end{picture}
\caption{The continuum image of IRAS 10214+4724 taken with the EVLA at 6.7 GHz. An emission peak (0.2 mJy beam$^{-1}$; 2.5\,$\sigma$) is found $\sim$\,4~arcsec from the position of the lens system (shown by the cross at 10$^{\rm h}$24$^{\rm m}$34.56$^{\rm s}$ $+$47$\degr$09$\arcmin$09.7$\arcsec$). Contours are shown at (-2, 2)~$\times$~0.08~mJy, the rms map noise. The restoring beam is shown as the grey ellipse in the lower left-hand side of the map.} 
\label{IRAS10214-cont}
\end{center}
\end{figure}

The EVLA continuum map of IRAS 10214+4724 at 6.7 GHz shows an emission peak of just 0.19~mJy~beam$^{-1}$ close to the expected position of the lens system, which is at the 2.5\,$\sigma$ level (see Fig.~\ref{IRAS10214-cont}). A low sub-mJy flux density for IRAS 10214+4724 at 6.7 GHz is consistent with the earlier observations of the system at cm-wavelengths \citep{rowan-robinson91,lawrence93}. The EVLA spectrum, extracted at the position of the three merging images is also presented in Fig.~\ref{IRAS10214-spectra}. We find a 2.5\,$\sigma$ peak in emission at 72\,$\pm$\,1~km\,s$^{-1}$, which is similar to the velocity of the tentative emission line found from the first epoch of Effelsberg observations. Unfortunately, the bandwidth of the EVLA data was not large enough to span the velocity of the other tentative detection made during the second epoch of Effelsberg observations. The candidate water line found using the EVLA has a peak flux density of 1.4 mJy,  a Gaussian FWHM of 3\,$\pm$\,1 km\,s$^{-1}$ and an integrated total intensity of 0.010~Jy~km\,s$^{-1}$. This corresponds to an isotropic luminosity of just 450~$L_\odot$ (for a lensing magnification of 50).

Our spectra of IRAS 10214+4724 that were taken with Effelsberg and the EVLA provide at best only tentative detections of water vapour emission at redshift 2.285, even though the candidate lines from Effelsberg are significant ($\sim$\,5\,$\sigma$ peaks). This is because we have not been able to verify these tentative lines from multiple observations with Effelsberg, or independently with the EVLA. It is certainly possible that the emission seen in the Effelsberg spectra is genuine, but is variable over the 4 month period between the observations. Water maser lines have been found to vary in flux density and line width over the time-scale of months before. For example, in the powerful FR\,II quasar 3C\,403, several maser lines at different velocities are seen to strongly vary over months to years \citep{tarchi07}. What is harder to explain is the variation seen within a few days during the second Effelsberg observation. However, given that the maser emission will come from a compact region, scintillation within our own galaxy (e.g. \citealt*{vlemmings07}) or microlensing within the lensing galaxy cannot be discounted. The EVLA spectrum shows an emission feature which is close in velocity to one of the tentative lines found with Effelsberg, but given the line's low significance (2.5\,$\sigma$) and width ($\sim$\,1 channel), it is not conclusive that this is a real detection. Note that the feature in the EVLA spectrum would be too weak to have been detected during the second epoch of Effelsberg observations. Therefore, we conclude that there is only a tentative detection of water vapour emission from IRAS 10214+4724.

\subsection{H~1413+117 -- The Cloverleaf}

The spectrum of the H~1413+117 lens system is presented in Fig.~\ref{all-spectra} and shows no evidence of water maser emission. There was significant local RFI detected with the Effelsberg radio telescope between 6.219 and 6.231 GHz, which resulted in part of the spectrum between 600 and 1200 km\,s$^{-1}$ from the systemic velocity of H~1413+117 having to be removed. The rms noise of the spectrum within a 10 km\,s$^{-1}$ wide channel is 0.9 mJy. This corresponds to an upper limit of $<$\,7700\,$L_\odot$ (3\,$\sigma$) for an unlensed water maser (assuming a lens magnification of $\sim$\,10). We note that our observations are about half as sensitive as those presented by \citet{wilner99}, but probe a much larger velocity range, out to around $\pm$\,2000\,km\,s$^{-1}$. The region of the Effelsberg spectrum affected by RFI was also not investigated by \citet{wilner99} because their spectrum was confined to velocities between $-$633 and $+$127\,km\,s$^{-1}$, relative to our definition of the systemic velocity of the quasar. Therefore, we also find no evidence of time-variable emission from this system.

\subsection{MS 1512--cB58}

The spectrum of MS 1512--cB58 is presented in Fig.~\ref{all-spectra} and shows no evidence of clear water maser emission within $\pm$\,2000\,km\,s$^{-1}$ from the systemic velocity. The rms noise of the spectrum is 1.2~mJy for a 10~km\,s$^{-1}$ wide channel. This corresponds to an upper limit of $<$\,2300\,$L_\odot$ (3\,$\sigma$) for the luminosity of any water maser emission within a 10~km\,s$^{-1}$ wide channel. This calculation assumes a lensing magnification of 50.

\subsection{SMM J16359+6612}

The spectrum of our final candidate water maser galaxy is also shown in Fig.~\ref{all-spectra}. We find no water maser emission in the spectrum, which has an rms noise of 0.9\,mJy within a 10\,km\,s$^{-1}$ wide channel. The upper-limit to the luminosity of water maser emission is therefore $<$\,1200\,$L_\odot$ (3\,$\sigma$) after correcting for the magnification from the cluster lens ($\sim$\,45). Comparing with the recent observations of SMM J16359+6612 by \citet{edmonds09} and \cite{wagg09}, we find that our spectrum is about half as sensitive as those taken with the EVLA. The Effelsberg observation confirms that there is no time-variable emission from this system and probes a much larger range in velocity. The EVLA observations were processed with the old VLA correlator and gave information between $-$406 and $-$106\,km\,s$^{-1}$, relative to our definition of the systemic velocity. 

\section{Discussion}
\label{discussion}

In this section, we discuss our results and present an analysis of the water maser luminosity function that is based on our survey of gravitationally lensed quasars and star-forming galaxies. We also discuss the prospects for finding non-lensed water maser galaxies in the distant Universe with the next generation of radio telescopes.

\subsection{Water masers at high redshift}

There have been several attempts to detect water vapour from gravitationally lensed galaxies. So far (including this work), eight gravitationally lensed quasars and star-forming galaxies at high redshift have been observed for 22.2 GHz water maser emission, with only one confirmed detection (MG J0414+0534 at redshift 2.64; \citealt{impellizzeri08}). We present in Table \ref{masers-past} the eight systems that have been observed and the 3\,$\sigma$ luminosity limits for those systems where no water masers were found. With the exception of APM 08279+5255 at redshift 3.911, all of the lens systems have similar redshifts, forming a sample of galaxies at redshift $\sim$\,2.5. Also, we find that for this sample, all of those systems with non-detections have luminosity limits which are below the measured luminosity of the water maser from MG J0414+0534. Therefore, we find the detection rate for powerful water masers with luminosities of $\sim$~10$^{4}$~$L_\odot$ from galaxies at redshift $\sim$\,2.5 to be one in seven.

We focussed our sample selection on those galaxies with evidence of large reservoirs of dust through previous detections of CO and/or from a high FIR-luminosity to investigate whether dusty galaxies were more likely to host powerful water maser systems. Note that the estimated cold-dust masses of the sample are of order $\sim$\,10$^8$~$M_\odot$ and that MG~J0414+0534 has a cold-dust mass of $\sim$\,10$^{7}$~$M_\odot$ \citep{barvainis02}. Our non-detections would suggest that the presence of dust is not enough to increase the likelihood of finding water masers at high redshift, in agreement with the findings of \cite{wagg09}. A correlation between dust and water maser emission has been tentatively found in the local Universe \citep{castangia08}; although, there is only one known local ultra-luminous infrared galaxy with water maser emission (UGC\,5101; \citealt{zhang06}). This correlation is strongest for water maser galaxies in the kilo-maser regime ($\leq 10$~$L_\odot$), that is, those which have some or most of their water maser emission associated with star formation \citep{surcis09}. Thus far, there is no evidence of such a correlation at moderate or high redshifts where galaxies are expected to have significantly more dust. CO may also not be a particularly good tracer of water maser emission. Although CO does trace the cold gas of a galaxy, and in the case of MG J0414+0534 the water maser emission and the CO have consistent velocities, CO tends to trace lower density gas (10$^4$ to 10$^6$~cm$^{-3}$) than is required for water maser emission (10$^7$ to 10$^{11}$~cm$^{-3}$). This implies that water maser emission and the overall dust and gas masses of a galaxy are not related in a simple way. This then leads to the possibility that jets may be a key ingredient for luminous maser emission at high redshift. Relativistic jets would provide a mechanism to chemically enrich the gas with water molecules through shocks, to collisionally excite the water molecules, as well as giving a strong continuum source to stimulate coherent emission. The next stage of our investigation is to carry out a larger survey of high-redshift AGN that are gravitationally lensed, in particular, focussing on those systems with known radio jets. From surveys at radio wavelengths, there are around 35 gravitationally lensed AGN with known relativistic jets (e.g. \citealt*{walsh79}; \citealt{lawrence84,winn00,browne03,myers03}).

\begin{table*}
\begin{center}
\caption{Summary of gravitationally lensed quasars and star-forming galaxies that have been searched for 22.2 GHz water masers. All luminosity limits for non-detections have been calculated at the 3\,$\sigma$ level and assume a 10 km\,s$^{-1}$ channel. Given the uncertainty in the magnification of any potential water maser systems, we also quote the magnification used for our calculations.}
\begin{tabular}{lcccrl} \hline
Lens system 		& Redshift	& Magnification		& Velocity coverage& \multicolumn{1}{c}{Luminosity}	& Reference			\\
       					&			&					& (km\,s$^{-1}$)		& \multicolumn{1}{c}{($L_\odot$)}	&						\\ \hline
MG J0414+0534	& 2.639		& 35				& 780				& $<$\,3300 						& \citet{wilner99}		\\
					&			&					& 1500				& 8300								& \citet{impellizzeri08}\\
APM 08279+5255	& 3.911		& 7					& 760				& $<$\,20000						& \citet{ivison06}		\\
RX J0911+0551	& 2.793		& 22				& 4200				& $<$\,7200						& this paper			\\
IRAS 10214+4724	& 2.285		& 50				& 500				& $<$\,980							& this paper			\\
SMM J14011+0252	& 2.565		& 4					& 300				& $<$\,2900						& \citet{wagg09}			\\
H1413+117			& 2.561		& 10				& 780				& $<$\,4600						& \citet{wilner99}		\\
					&			&					& 4000				& $<$\,7700						& this paper			\\
MS1512$-$cB58	& 2.720		& 50				& 4200				& $<$\,2300						& this paper			\\
SMM J16359+6612	& 2.516		& 45				& 600				& $<$\,1800						& \citet{edmonds09}	\\
					&			&					& 300				& $<$\,650							& \citet{wagg09}		\\			
					&			&					& 3600				& $<$\,1700						& this paper			\\
\hline
\end{tabular}
\label{masers-past}
\end{center}
\end{table*}

\subsection{Evolution of the water maser luminosity function}
\label{evo}

The water maser luminosity function describes the co-moving number density of water maser galaxies with luminosity, $L_{\rm H_2O}$, per logarithmic interval in $L_{\rm H_2O}$ and can be used to investigate the density and luminosity evolution of water maser galaxies as a function of redshift. The water maser luminosity function for the local Universe was calculated using up to 78 galaxies at distances less than 235~Mpc by \citet{henkel05} and \citet{bennert09}, and was found to be $\Phi (L) \propto L^{-1.4\pm0.1}_{\rm H_2O}$. \citet{bennert09} also considered the detection of SDSS J080430.99+360718.1 at redshift 0.66, and found that, given the sensitivity of their survey of 274 Seyfert 2 galaxies with the Green Bank and Effelsberg telescopes, the detection was consistent with the local water maser luminosity function. 

We have investigated whether the detection of a water maser with an unlensed luminosity of 8300~$L_\odot$ from MG J0414+0534 at redshift 2.64 is consistent with the water maser luminosity function found at low and moderate redshifts (see \citealt{impellizzeri08}). The probability of detecting the water maser from MG J0414+0534 was estimated by integrating the luminosity function of \citet{henkel05}, with a slope of $-$1.4, over all luminosities greater than 10~$L_\odot$. It was assumed that only those water masers with luminosities stronger than 10~$L_\odot$ are associated with AGN, which is the case in the local Universe. However, this split in luminosity may not be so clear at higher redshifts where it could be harder to separate the water maser emission associated with either star-formation or AGN activity. We found that for galaxies with water masers, the fraction that have maser emission greater than 8300~$L_\odot$ is predicted to be just 8\,$\times$\,10$^{-5}$. This would imply that to find such a water maser system from a single pointing is rare, and would require a large volume to be surveyed for a detection to be made. However, we have carried out a targeted survey of AGN and star-forming galaxies. In the local Universe, the detection rate of water maser galaxies is approximately a few per cent \citep{braatz07}. The combined probability of observing an AGN and finding an 8300~$L_\odot$ water maser galaxy is estimated to be just $\sim$10$^{-6}$. Therefore, our detection of at least one water maser galaxy at redshift $\sim$\,2.5 implies that there has to be some change in the water maser luminosity function from what is found out to moderate redshifts ($z\sim$~0.7). The natural consequence of this evolution is that powerful water maser galaxies must be significantly more abundant at redshift $\sim$\,2.5 when compared to the number found locally. As we have only a single detection at high redshift it is not possible to place any strong constraints on what form this change in the luminosity function could be. For example, there could be a change in the slope of the luminosity function, or strong density and/or luminosity evolution.

\subsection{Predictions for the SKA and pathfinders}

The next generation of radio telescopes, such as the EVLA, MeerKAT (Karoo Array Telescope) and the SKA (Square Kilometre Array), will allow deep and/or wide-field surveys to be carried out at frequencies where water maser galaxies can be found at cosmologically interesting epochs ($z >$~1). To investigate the feasibility of blind surveys with these telescopes, we have calculated the number of water maser galaxies we would expect to find.  We give the expected specifications of the EVLA, MeerKAT and the SKA in Table \ref{surveystats} that were used for our calculations.

We predicted the number of water maser galaxies by integrating the local water maser luminosity function of \citet{henkel05},
\begin{equation}
\Phi = (1 \times 10^{-3})~L_{\rm H_2O}^{-1.4}~\rm{Mpc^{-3}~(0.5~dex)^{-1}}
\end{equation}
(where $L_{\rm H_2O}$ is expressed in solar luminosities) between 10~$L_\odot$~$\leq L_{\rm H_2O} \leq$~10$^{4.66}$~$L_\odot$ (i.e. twice the most luminous water maser known) and over the volume of sky that can be surveyed (defined by the field-of-view, frequency coverage and total instantaneous bandwidth). We assume that the luminosity function is a power-law without a knee before our luminosity cut-off.  We also parameterize any evolution of the luminosity function with redshift using,
\begin{equation}
(1+z)^m
\end{equation}
where $m$ is 0, 4, and 8 for no, moderate and extreme evolution, respectively, up to redshift 2.2, after which there is assumed to be constant density evolution.

How far we can probe down the luminosity function at different redshifts is dependent on our detection limit. This is calculated by first determining the system equivalent flux density (SEFD) using,
\begin{equation}
{\rm SEFD} = 2~k~T_{\rm sys} / A_{\rm eff}
\end{equation}
where $k$ is the Boltzmann constant and $A_{\rm eff} / T_{\rm sys}$ is given in Table~\ref{surveystats}. The limiting flux-density for the detection of a maser line is calculated using,
\begin{equation}
S_{\rm rms} = \frac{1}{\eta_c }~{\frac{\rm SEFD}{\sqrt{n_{\rm pol}~\Delta\nu~t}}}
\end{equation}
where $\eta_c$ is the correlator efficiency and is taken as 0.7, $n_{\rm pol}$ is the number of polarizations, $\Delta\nu$ is the line width and $t$ is the observing time. For our calculations, we assume a line width of 10~km\,s$^{-1}$ and dual polarization observations. The luminosity limit is then found from Equation~\ref{lum}.

We show the results in Table~\ref{predictions} of our simulations for a single pointing, 100 pointings and quarter-sky surveys (where appropriate) for the EVLA, MeerKAT and the SKA. We considered total integration times of 3000~h on-source to complete the surveys. The time per pointing is simply the total time divided by the number of pointings and frequency set-ups needed to cover the full range in frequency. We show the number of detections that are expected for the three evolutionary scenarios in Table~\ref{predictions}, but here we only discuss the results for moderate density evolution, $(1+z)^4$. 

\begin{figure}
\begin{center}
\setlength{\unitlength}{1cm}
\begin{picture}(6,7.8)
\put(-3.5,8.6){\includegraphics{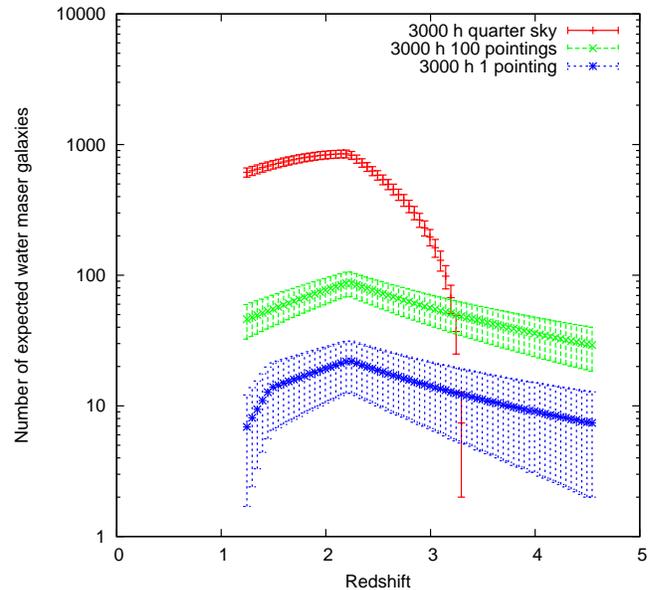}}
\end{picture}
\caption{The number of water maser galaxies that are expected to be found (3\,$\sigma$ level) with the SKA Phase 2. In total, 3000 h is used for either a single pointing, 100 pointings or quarter sky surveys. These calculations assume that the form of the luminosity function remains unchanged from that derived locally and the population of galaxies that host water masers evolve with $(1+z)^4$, up to redshift 2.2., after which there is constant density evolution. The error bars are at $\pm$\,2\,$\sigma$ confidence level.} 
\label{ska-plot}
\end{center}
\end{figure}

Taking the EVLA and MeerKAT results first, we find that these two instruments are not particularly useful for blind surveys of high redshift water maser galaxies, given their limited sensitivity and fields-of-view. However, the sensitivities of these telescopes are good enough to detect water masers out to redshift 1--2 with $\sim$~1000--5500~$L_{\odot}$  using 15--30~h on-source (i.e. unlensed water masers with luminosities similar to or lower than that of MG J0414+0534). Therefore, the EVLA and MeerKAT will be important for targeted searches of candidate water maser galaxies in the future, similar to those carried out here for the gravitationally lensed water maser project. We find that the proposed sensitivity and field-of-view of the SKA Phase 2 will be sufficient to detect many water maser galaxies out to high redshifts. In Fig.~\ref{ska-plot} the number of predicted detections as a function of redshift for a single pointing, 100 pointings and quarter-sky surveys with the SKA are shown. The quarter-sky survey is predicted to detect at least 23500 water maser galaxies out to around redshift 3, beyond which the sensitivity is not good enough to make detections in meaningful numbers. The time on-source per pointing for this wide-field survey is around 5~s. The 100 and single pointing surveys are expected to have the sensitivity to detect around 3700 and 900 water maser galaxies, respectively, out to redshift 4.56, the redshift cut-off of our proposed surveys.

These predictions are based on a number of assumptions, which we now discuss. First, we assume that the luminosity function of water maser galaxies is consistent with that derived locally by \citet{henkel05} and \citet{bennert09}. This luminosity function is based on only 78 nearby galaxies, and does not probe the high luminosity regime ($>$~10$^4$~$L_{\odot}$). We have adopted a high luminosity cut-off at 10$^{4.66}~L_\odot$ for making our predictions, but the resulting number of detections is not particularly sensitive to the choice of this value. There is evidence for a knee in the \citet{henkel05} luminosity function around 10$^3$~$L_{\odot}$. However, this is thought to be due to there being few high luminosity masers in the limited volume surveyed, as opposed to a genuine steepening of the spatial distribution of high luminosity water masers. We have tried to limit the effect of a potential cut-off in the luminosity function by only integrating the luminosity function up to twice the luminosity of the most luminous water maser known. Also, from the detection of a luminous water maser system at redshift 2.64, it was found that there must be some change in the local luminosity function with redshift. Therefore, our predictions must give a lower limit to the number of water masers that we would expect to find. The second major assumption in our predictions is that the number of potential host galaxies with the physical conditions needed for water maser emission will increase with $(1+z)^4$, that is, similar to the merger-rate of galaxies found from optical surveys \citep{lefevre00,ryan08}. We have tried to account for this uncertainty by using no and extreme evolution in the luminosity function. Although no change in the number density of water maser galaxies with redshift is unlikely, extreme evolution may be possible. For example, in the case of optically bright quasars the evolution goes as $(1+z)^6$ up to redshift $\sim$\,2.5 (\citealt{briggs98}; \citealt*{hewett93}; \citealt{schmidt95}).

\begin{table*}
\begin{center}
\caption{The specifications for the different arrays used to calculate the expected number of water maser galaxies from $z=$~0.5 up to $z=$~4.5 that can be found from blind surveys.}
\begin{tabular}{lllccccc} \hline
Telescope		& Frequency	& Number of	& Antenna diameter	& Aperture efficiency	& $T_{\rm sys}$	& Total simultaneous bandwidth	& $A_{\rm eff} / T_{\rm sys}$\\
				& (GHz)		& antennas	& (m)			&				& (K)				& (GHz)						& (m$^2$ / K)	 		\\\hline
EVLA			& 4--12		& 27			& 25				& 0.70			& 38				& 2.0							&   245				\\
MeerKAT Phase 4	& 10--14		& 80			& 12				& 0.70			& 30				& 4.0							&   210				\\
SKA Phase 2		& 4--10		& 2000		& 15				& 0.70			& 30				& 2.0							& 8250				\\
\hline
\end{tabular}
\label{surveystats}
\end{center}
\end{table*}

\begin{table*}
\begin{center}
\caption{The predicted number of water maser galaxies found from blind surveys. The sky-areas are quoted at 10~GHz, except for the EVLA observations between 4 and 8 GHz, which are defined at 5 GHz. The luminosity limits are for 10~km\,s$^{-1}$ channels within the range of redshifts that are observable. The number of detections are for different evolutionary models given by $(1+z)^m$, where $m$ is 0, 4, and 8 for no, moderate and extreme evolution, respectively. The $L_{3\,\sigma}$ denotes the 3\,$\sigma$ limiting luminosity that can be detected across the full frequency range for the given time on-source per pointing.}
\begin{tabular}{lllllllllll} \hline
Telescope		& Frequency 	& Redshift	& Total time	& Pointings				& Number of 			& Sky area					& $L_{3\,\sigma}$	& \multicolumn{3}{c}{Detections}	\\
				& (GHz)	  	&				& (h)			&							& frequency settings	& (sr) 						& ($L_{\odot}$)	& $m =$~0 	& $m =$~4		& $m =$~8 	\\ \hline
EVLA			& 8--12		 &	0.85--1.78	& 3000			& 1							& 2						& 1.7\,$\times$\,10$^{-6}$	& 90--450			& 0				& 0				& 30			\\
				&			   &				& 3000			& 100						& 2						& 1.7\,$\times$\,10$^{-4}$	& 900-4500			& 0				& 0				& 100			\\
				& 4--8		   &	1.78--4.56	& 3000			& 1							& 2						& 6.7\,$\times$\,10$^{-6}$	& 450--3100		& 0				& 0				& 250			\\
				&			   &				& 3000			& 100						& 2						& 6.7\,$\times$\,10$^{-4}$	& 4500--31000		& 0				& 0				& 850			\\
MeerKAT Phase 4	& 10--14	   &	0.59--1.22	& 3000			& 1							& 1						& 7.3\,$\times$\,10$^{-6}$	& 30--160			& 0				& 0				& 30			\\		
				&			   &				& 3000			& 100						& 1						& 7.3\,$\times$\,10$^{-4}$	& 300--1600		& 0				& 10			& 130			\\
SKA Phase 2		& 4--10		   &	1.22--4.56	& 3000			& 1							& 3						& 4.7\,$\times$\,10$^{-6}$	& 7--110			& 10			& 900	 		& 82000		\\
				&			   &				& 3000			& 100						& 3						& 4.7\,$\times$\,10$^{-4}$	& 70--1100			& 50			& 3700			& 330000		\\
				&			   &				& 3000			&  6.7\,$\times$\,10$^{5}$	& 3						& 3.14						& 5600--90000		& 400			& 23500 		& 1800000		\\		
\hline
\end{tabular}
\label{predictions}
\end{center}
\end{table*}

\section{Summary}
\label{conclusion}

We have presented new radio spectra for five gravitationally lensed quasars and star-forming galaxies at redshifts between 2.29 and 2.80. We searched these galaxies for 22 GHz (rest frame) water masers that potentially could be used to study the properties of their central supermassive black hole. No new water masers were found, although, we do have a tentative detection of water vapour from the quasar IRAS 10214+4724 at redshift 2.285. Our survey has now searched for water masers in six gravitationally lensed quasars and star-forming galaxies, doubling the number of high redshift lensed galaxies surveyed so far, and finding the only known high redshift water maser system \citep{impellizzeri08}.

We have presented an analysis of the high redshift water maser luminosity function that is based on the results from this paper and from \citet{impellizzeri08}. We again confirm that there must be some evolution in the water maser luminosity function to explain the detection of the luminous (8300~$L_{\odot}$) water maser system from MG J0414+0534 at redshift 2.64. We therefore conclude that water masers are more abundant at higher redshifts than found locally, and surveys with the next generation of radio arrays have the potential to find many more water maser systems in the future. However, blind surveys for water maser galaxies will only be possible with a sensitivity like that of the proposed SKA (Phase 2).

\section*{Acknowledgments}

Our results are based on observations with the 100 m telescope of the MPIfR (Max-Planck-Institut f{\" u}r Radioastronomie) at Effelsberg and the EVLA which is operated by the National Radio Astronomy Observatory and is a facility of the National Science Foundation operated under cooperative agreement by Associated Universities, Inc. OW was supported by the Emmy-Noether-Programme of the Deutche Forschungsgemeinschaft under reference WU 588/1-1.

\bsp

\label{lastpage}

\end{document}